# Ultra-High-density 3D vertical RRAM with stacked JunctionLess nanowires for In-Memory-Computing applications

M. Ezzadeen, D. Bosch, B.Giraud, S. Barraud, J.-P. Noël, D. Lattard, J. Lacord, J.-M. Portal, F.Andrieu

*Abstract*— The Von-Neumann bottleneck is a clear limitation for data-intensive applications, bringing In-Memory Computing (IMC) solutions to the fore. Since large data set are usually stored in Non-Volatile Memory (NVM), various solutions have been proposed based on emerging memories, such as OxRAM, that rely mainly on area hungry, one transistor (1T) one OxRAM (1R) bit-cell. To tackle this area issue, whereas keeping the programming control provided by 1T1R bit-cell, we propose to combine Gate-All-Around stacked junctionless nanowires (1JL), and OxRAM (1R) technology to create a 3D memory pillar with ultra-high-density. Nanowire junctionless transistors have been fabricated, characterized, and simulated to define current conditions for the whole pillar. Finally, based on SPICE simulations, we demonstrated IMC scouting logic operations up to three pillar's layers, with one operand per layer.

## I. INTRODUCTION

With the increasing number of connected devices and Artificial Intelligence (AI) applications, the "data deluge" is a reality, making energy efficient computing systems a must have. Unfortunately, the well-known Von-Neumann architecture is computing centric and not data-centric. Actually, data movements in the memory hierarchy result in 50% energy waste [1]. To overcome this limitation, In/Near-Memory Computing (IMC/NMC) rises to be a solution with the co-location of data and logic operations, reducing drastically data movements.

Several IMC approaches can be found in literature, shared between volatile (DRAM or SRAM) and non-volatile memory (Resistive memories as well as charge storage) [2]. Since massive data are mainly stored in Non-Volatile Memory (NVM), they are naturally a good candidate for low-power IMC. In the landscape of the NVM, the RRAM are a promising solution, since they can be scaled down to 10×10nm² with a good reliability and low voltage operation [3]. In this context, numerous boolean IMC/NMC approaches based on RRAM have been proposed like [4-7]. Mainly, two concepts are competing: the first one implies read and write operation on the RRAM [4-5] and the second one implies only read operations [6-7], preserving in this way the energy budget and the memory endurance. This last concept is named Pinatubo or SCouting Logic (SCL).

Regarding RRAM, one flavor, namely Oxide-based RAM (OxRAM), presents very interesting features like low programming voltage compared to Flash memory, fast switching time and a very friendly integration with CMOS material. The most aggressive density might be reached considering back-end selector (1S) coupled with one OxRAM (1R). However, this 1S-1R configuration in crossbar array with really reduced read margin, is hardly compatible with IMC, since sneak path current of unselected cells may strongly limit the array size considered for an IMC operation [8]. Thus the majority of silicon-proven OxRAM circuits are based on the well-known, one access transistor (1T) coupled with one OxRAM (1R). The 1T-1R configuration allows controlling low sneak current of unselected cell as well as the current compliance of the selected OxRAM enabling IMC solution as proposed in [9].

The main drawback is a high silicon surface occupation due to the select transistor. To overcome this density issue, one can think going 3D by stacking 1T/1S-1R on top of the others. For instance, 8 layers vertical self-selective RRAM are proposed in [10], but this solution is only lightly suitable for large-scale IMC, since it does not co-integrate one transistor with each RRAM. A boolean IMC approach has been reported in [11] on a Vertical RRAM Pillars with a 1T-4R configuration. In this case, boolean operations takes many steps and induce several write operations, which can be detrimental for OxRAMs endurance. Here also, one transistor is not associated with each OxRAM.

Actually, fine grain transistors stacking is a challenge due to the thermal budget restrictions for top-tier sequential integration [12]. At the same time, for sub-7nm nodes, Gate-All-Around (GAA) Stacked nanowires with an excellent electrostatic control have been demonstrated, improving both performance and density [13]. As demonstrated in [14], a high number of stacked Si channels can be used and laterally co-integrated with OxRAM, paving the way for 3D ultra-high density integration (Fig. 1). In addition, uniformly doped channel without junctions (junctionless) devices have emerged as an alternative to conventional devices due to their ease of fabrication and a higher gate oxide reliability [15].

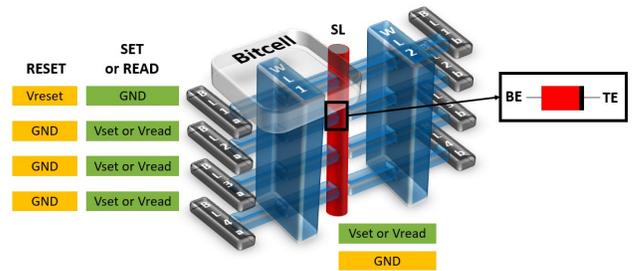

Fig. 1: 3D pillar structure scheme with 4 layers, including BL, WL and SL voltages configurations to program (SET/RESET) and read a given bitcell.

In this context, the aim of this article is to demonstrate SCL operations, based on our new one junctionless nanowire (1JN) with one lateral OxRAM (1R) ultra-dense pillar structure presented in section II. In section III, the fabrication and electrical characterization of junctionless transistor is described and TCAD simulation is performed, to define the transistor configuration compatible with the

This work was funded by French Public Authorities through the NANO2022, LabEx Minos ANR-10-LABX-55-01 and by the European Research Council (ERC) through My-CUBE project.

M. E., D. B, B. G., S. B., J.-P. N., D. L., J. L. and F. A are with CEA-LETI/LIST, Univ. Grenoble Alpes, 17 rue des Martyrs, 38054 Grenoble, France.

J.-M. P. and M. E. are with Aix-Marseille Université, IM2NP, CNRS UMR 7334, F-13453 Marseille, France.

different OxRAM programming scenarios. The SCL operations in our new pillar structure, is validated with SPICE simulations based on these technological inputs in section IV. Finally, section V concludes the paper.

## II. A New 3D Memory: pillar structure and memory operation

The memory technology we consider in this paper mixes at a fine grain level the most advanced and energy-efficient CMOS technology (GAA stacked nanowires) with a lateral OxRAM. The bitcell topology is represented in Fig.1. Conversely to GAA transistors, each source and drain of the stacked nanowires is independent. Thus each nanowire can address an OxRAM, whose materials (oxide and top electrodes) are deposited in a vertical pillar; the bottom electrode of the memory being localized at the drain side of each transistor. So, the final structure includes two times $n$ stacked nanowires with a common gate (WL1, WL2), separate drains (BL1a to BL$n$a, and BL1b to BL$n$b) and a common pillar called SL gathering the sources.

Programming and read operations on the pillar are performed in a classical manner, like in standard 1T1R memories. SET, RESET or READ voltages are applied on BLs or SLs. The bitcells of the same column which are unused are short circuited, while access transistors of unused columns are off.

Such a memory technology presents the advantages of 1T1R structures with a high integration provided by GAA transistor technology, as long as the transistor features a thin gate oxide (GO1 transistor). This may be possible since GO1 devices already proved their compatibility with OxRAM endurance requirements [16]. In our simulations, four stacks (n=4) are considered but up to seven ones were already demonstrated experimentally [14]. Moreover, in order to extend the up-scaling of such a technology in the vertical dimension, junctionless nanowire transistors are studied in section III. Actually, the integration of junctionless transistor relaxes the constraints in terms of source/drain doping for multiple stacked nanowires (n>4).

## III. Analysis of JunctionLess Devices

### A. JunctionLess process flow and performances

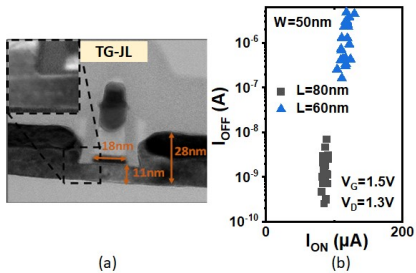

Fig. 2: (a) TG-JL TEM cross-section of JAM (N+-N-N+) transistors. (b) $I_{OFF}$-$I_{ON}$ for W=50nm and L=60nm and 80nm.

To provide insights on the pillar performances, especially on the current driven by junctionless (JL) transistors, a Tri-Gate-JL (TG-JL) NMOS were fabricated. The Si thickness of TG-JL transistors is 11nm with a channel doping of $7\times10^{18}$ at/cm$^3$. The gate stack consists of HfO$_2$ dielectrics (equivalent oxide thickness EOT=1nm), 5nm ALD TiN, and polysilicon layers. After the gate patterning, a 8nm SiN spacer is defined. To lower the source/drain access resistance the source and drain are highly doped by ion implantation. The process flow is described in [17] and a device cross-section is given in Fig. 2-(a). To enable read and write operation of RRAM memory, a high drive current is required to set a logic state into the memory element, together with a small ON variability to reduce the resistance distribution. That's why the drive current and variability of TG-JL were experimentally extracted.

First, the tradeoff between the ON- and OFF-state currents ($I_{ON}$-$I_{OFF}$, see Fig. 2-(b)) shows that in average, devices of 50nm width and 80nm (respectively 60nm) gate length drive 87µA (120µA) ON-current for an OFF current of $10^{-9}$A ($10^{-6}$A) at $V_D$=1.3V and $V_G$=1.5V.

Secondly, as far as global variability is concerned, NMOS shows a threshold voltage $V_T$ (extracted at constant current) of 0.18V and a $V_T$-deviation of 48mV at W=50nm and L=80nm (Fig. 3). It should be pointed out that the higher sub-threshold variability reported for junctionless transistor vs. inversion-mode ones does not translate necessary into a higher ON current variability thanks to screening effects at high gate voltage [17]. From the measured Pelgrom plot (Fig. 4-(a)), we extrapolated a 42mV variation reduction by enlarging the device width to W=75nm. In the next part, the electrostatics and drive current capability for W=75nm Gate-All-Around transistors will be explored with TCAD simulations.

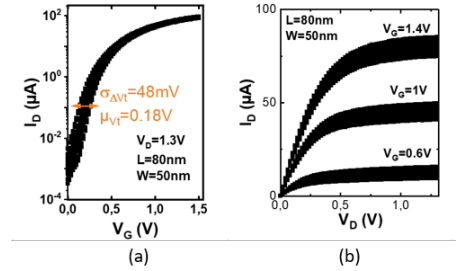

Fig. 3: (a) $I_D$-$V_G$ curves for W=50nm and L=80nm junctionless nMOS at $V_D$=1.3V. (b) $I_D$-$V_D$ curves for W=50nm and L=80nm.

### B. Drive current for stacked nanowires (W=75nm)

To have insights on JL drive current in the memory array, TCAD simulations (Fig. 4-(b)) have been performed considering TG-JL and GAA nanowire (GAA-NW) configurations at W=75nm. Sentaurus Device tool from Synopsys was used for this study. Compared to the TG-JL (REF in Table I) at W=50nm, simulated TG-JL at W=75nm drives 50% more current at a slightly higher OFF current. Going to a JL-GAA-NW configuration increases both electrostatic control and drive current (-3 decades on $I_{OFF}$ and +70% on $I_{ON}$). Finally, increasing the channel doping $N_D$ (from $7\times10^{18}$ at/cm$^3$ to $10^{19}$at/cm$^3$) enables to increase by +150% the drive current compared to the experimental Ref at the same $I_{OFF}$.

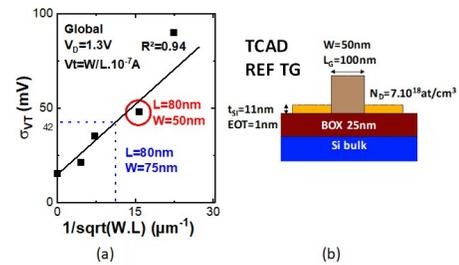

Fig. 4: (a) Experimental pelgrom plot measured for TG-JL nMOS (global variability). (b) REF structure for TCAD simulation with the defined process parameters.

In order to assess the impact of the current delivered by the JL transistors configurations, four SET conditions were considered in the following: weak (70µA), Light typical (100µA), Strong Typical (150µA) and Strong (200µA) conditions.

TABLE I
RECAP TABLE OF TCAD RESULTS

| Configuration | TG | TG | GAA | GAA |
|---|---|---|---|---|
| W (nm) | 50 | 75 | 75 | 75 |
| $N_D$ (at/cm³) | $7.10^{18}$ | $7.10^{18}$ | $7.10^{18}$ | $10^{19}$ |
| $\log(I_{OFF})$ (A) | -7 | -6.5 | -10 | -7 |
| $I_D$ @ $V_D$=1.3V, $V_G$=1.5V (µA) | 50 | 75 | 86 | 126 |

### C. OxRAM distribution extraction

We fabricated OxRAM cells of 10nm $HfO_2$/Ti 10nm/TIN stack deposition on top of a TiN bottom electrode and arranged into 4kbits 1T1R array (details are given in [18]). Resistance distributions (mean µ and standard deviation σ) for previously defined current compliance are extracted for a $T_{pulse}$=100ns pulse width and a 2V source line voltage (table II and illustrated in Fig. 8). The RESET conditions for $V_{bl,reset}$=2.5V BL voltage and $T_{pulse}$=100ns corresponds to a lognormal HRS distribution with parameters µ=120kΩ and σ=0.63.

TABLE II
SET CONDITIONS

| SET condition | Compliance current (µA) | Resistance parameters µ(kΩ) /σ(kΩ) |
|---|---|---|
| Strong | 200 | 5.2/0.58 |
| Strong Typical | 150 | 5.7/0.73 |
| Light Typical | 100 | 8/1.3 |
| Weak | 70 | 10/2 |

## IV. SCOUTING LOGIC IN THE 1JN-1R ULTRA DENSE PILLAR

In this section, the feasibility of the scouting logic on our 1JN-1R new pillar structure is demonstrated with electrical simulations based on technological inputs. In this aim, 4 SET programming conditions are considered to increase the numbers of functional layers and thus the number of operands processed in parallel.

All the SPICE simulations presented in this section use (1) an OxRAM model based on experimental distributions given in Table II and (2) a transistor model from a commercial design kit emulating the performances of the junctionless nanowire given Table I. Variability is considered up to 6 sigma, and Monte Carlo simulations are performed with 1000 runs.

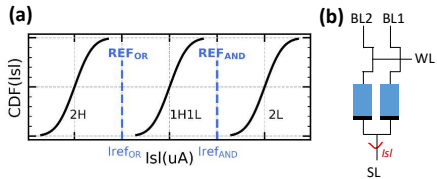

Fig. 5: SCL principle: depending on the RAMs states the current on SL exhibits different values comparable to the current reference of various logic operation.

The SCL principle is first re-called and illustrated considering 2 active layers, with the current distribution and the current reference of the Boolean operation (Fig. 5-(a)) and the circuit equivalent scheme (Fig. 5-(b)). As reported in [7] LRS state corresponds to a logic state '1', and HRS to a logic state '0'. When two layers are simultaneously activated, the corresponding OxRAM are subjected to the same read voltage applied between SL and the corresponding BL (BL1 and BL2). Depending on the resistance values of the two OxRAM (2 HRS, 1 HRS and 1 LRS, 2 LRS), the total current flowing through the line SL will take different values belonging to one of the distributions, as depicted Fig. 5-(a). A current reference (Iref) is then chosen between each current distribution. Boolean operations - AND, OR, XOR - are then simply achieved by sensing the SL current and comparing it to the appropriate reference(s). For instance, to perform an "OR" operation, the SL current Isl is compared to the leftmost reference in Fig. 5, $Iref_{OR}$: if Isl is below $Iref_{OR}$, it means that the two accessed memristors are in HRS state (i.e. both represents a logic state '0'), and the "OR" output is therefore '0'; if Iread is above $Iref_{OR}$, at least one of the memristors is in LRS (logic state '1'), so the "OR" output is '1'. XOR operation is performed by combining the results of the OR and the AND operations. This approach can be extended to n operands, with the simultaneous activation of n layers.

To implement SCouting Logic (SCL) in the pillar, BLs corresponding to the activated layers are grounded, whereas a read voltage is applied to SL. Like for the usual READ operation (see Fig. 1), un-selected bitcells of the selected WL are inhibited ($V_{BL}$ = Vread), and un-selected WL have their access transistor gate grounded ($V_{WL}$ = 0 V).

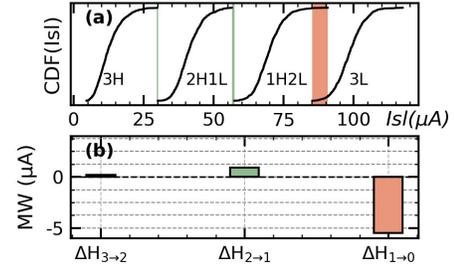

Fig. 6: Scouting logic results with Light Typical SET on three layers represented according to (a) current distributions and (b) Memory Windows values between current distributions.

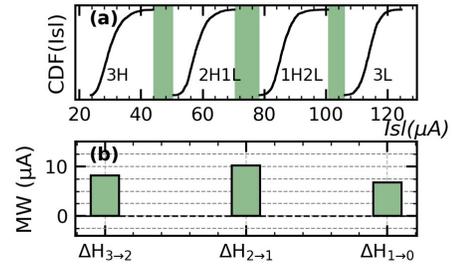

Fig. 7: Scouting logic results with Strong SET on three layers represented according to (a) current distributions and (b) Memory Windows values between current distributions.

To implement SCL successfully on a given number of layers, current distributions corresponding to the different combinations of HRS and LRS states must not overlap. Fig. 6-(a) shows the current distributions when 3 layers are activated and a Light Typical SET programming condition is considered. We observe that the third distribution (one HRS and two LRS cells) and the fourth one (three LRS cells) overlaps, whereas a slightly positive Memory Windows

(MW) is existing for the other cases. Fig. 6-(b) gives for each pair of side distributions the MW value as a positive value, whereas the overlap value is given as a negative value, both expressed in in µA. The same results are presented in Fig. 7, but with Strong SETs. We notice that all MW are preserved, making the implementation of SCL possible.

SCL performance is linked to the number of operands processed in parallel, i.e. the number of layers successfully accessed in parallel without overlap. Thus, a successful operation highly depends on HRS and LRS distributions, and therefore on SET and RESET conditions. However, LRS distributions seems to have a bigger impact on current distributions overlapping [19], as LRS cells conductance is higher. Fig. 8 shows MW and overlaps for various SET conditions, from one (Fig. 8-(a)) to four (Fig. 8-(d)) activated layers, with $V_{SL}$= 0.5 V. We notice that, (1) classical read operations (only 1 layer activated) can be achieved by all SET conditions (Fig. 8-(a)), (2) scouting logic can be performed with up to three layers with the two strongest SET conditions, and (3) as expected, MW are higher for stronger SET conditions.

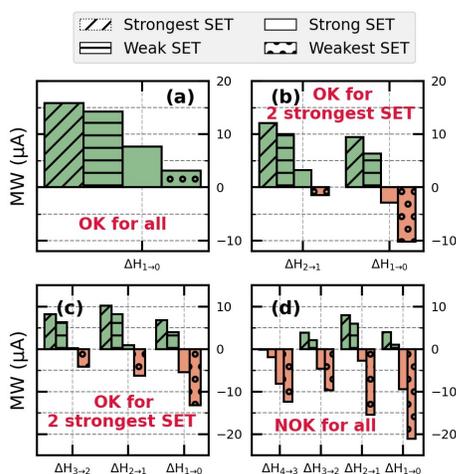

Fig. 8: MW and overlap as a function of SET conditions, on a pillar of four layers, with one (a) to four (d) activated layers @ Vsl=0.5V.

## V. CONCLUSION

We proposed a high-density 3D vertical RRAM pillar with stacked Junctionless nanowires (1JN-1R configuration) for In-Memory-Computing purposes. Based on electrical characterization and TCAD/SPICE simulations, we demonstrated the capability of junctionless transistors for performing classical read/write operations, and for delivering sufficient compliance current during SET, enabling SCL with up to three operands. Future work will consider simulations on a whole memory cube with the corresponding periphery to demonstrate the concept on a larger scale.